\documentclass[pre,twocolumn,longbibliography,amsmath,amssymb,floatfix]{revtex4-1}

\usepackage{graphicx}
\usepackage{dcolumn}
\usepackage{bm}
\usepackage{soul}
\usepackage[usenames,dvipsnames]{xcolor}
\usepackage{mciteplus}
\usepackage{braket}

\newcommand{\bnabla}{{\bm \nabla}}
\newcommand{\uref}{{\bm u}_{\rm ref}}
\newcommand{\bu}{{\bm u}}
\newcommand{\bx}{{\bm x}}
\newcommand{\bk}{{\bm k}}

\begin{document}
\title{Synchronization to big-data:\\nudging the
Navier-Stokes equations for data assimilation of turbulent flows}
\author{Patricio Clark Di Leoni$^{1,2} $\email{patricio.clark@roma2.infn.it},
        Andrea Mazzino$^3$\email{andrea.mazzino@unige.it} and 
        Luca Biferale$^1$\email{luca.biferale@roma2.infn.it}}
\affiliation{$^1$Department of Physics and INFN, University of Rome Tor
Vergata, Via della Ricerca Scientifica 1, 00133 Rome, Italy.\\
$^2$ Department of Mechanical Engineering, Johns Hopkins University, Baltimore, Maryland 21218, USA. \\
$^3$Department of Civil, Chemical, and Environmental Engineering and
INFN, University of Genova, Genova 16145, Italy.}
\date{\today}

\begin{abstract}
Nudging is an important data assimilation technique where partial field
measurements are used to control the evolution of a dynamical system
and/or to reconstruct the entire phase-space configuration of the
supplied flow. Here, we apply it to the toughest
problem in fluid dynamics: three dimensional homogeneous and isotropic
turbulence. By doing numerical experiments we perform a systematic
assessment of how well the technique reconstructs large- and
small-scales features of the flow with respect to the quantity and the
quality/type of data supplied to it. The types of data used are: (i)
field values on a fixed number of spatial locations (Eulerian nudging),
(ii) Fourier coefficients of the fields on a fixed range of wavenumbers
(Fourier nudging), or (iii) field values along a set of moving probes
inside the flow (Lagrangian nudging).  We present state-of-the-art
quantitative measurements of the scale-by-scale  {\it transition to  synchronization}  and a
detailed discussion of the probability distribution function of the
reconstruction error, by comparing  the nudged field and the
{\it truth} point-by-point.  Furthermore,  we show that for more complex
flow configurations, like the case of anisotropic rotating turbulence,
the presence of cyclonic and anticyclonic structures leads to
unexpectedly better performances of the algorithm.  We discuss potential
further applications of nudging to a series of applied flow
configurations, including the problem of field-reconstruction in thermal
Rayleigh-B\'enard convection and in magnetohydrodynamics (MHD), and to
the determination of optimal parametrisation for small-scale turbulent
modeling. Our study fixes the standard requirements for future
applications of nudging to complex turbulent flows.
\end{abstract}
\maketitle

\section{Introduction} 

Turbulence is the chaotic, non-linear, and multiscale motion observed in
fluids. From astro- and geo-physical flows to engineering ones, it is a
problem that surrounds us all \cite{Davidson,Pope}. Thus, observing,
measuring, reconstructing, and then predicting the evolution of
turbulent flows are highly important tasks with direct consequences to
our day to day lives. A paradigmatic example is given by the problem of
state estimation in geo-sciences, of particular importance to numerical
weather prediction (NWP) \cite{Kalnay}.  The chaotic and multiscale
nature of turbulence makes these tasks very difficult, as any small
error in the initial conditions will make predictions diverge from the
truth and as it is not easy to access all active modes in a fluid flow.
This is particularly troublesome when one considers that in a turbulent
flow the number of active degrees of freedom (dof) grows with the
Reynolds number as $\#_{dof} \propto Re^{9/4}$, with $Re = UL/\nu$,
given in terms of the typical rms velocity $U$, the energy containing
scale $L$ and the fluid viscosity, $\nu$.  Data assimilation (DA) is the
family of mathematical protocols  used to reconstruct the initial state
of a dynamical system, out of a series of previous partial measurements,
in order to ensure that any future predictions will be as faithful as
possible to what the actual physical reality will be, and has proven to
be of key importance in the development of modern NWP
\cite{Bannister16,Carrassi18, Bauer15}.

\begin{figure}[h]
    \centering
    \includegraphics[width=8.cm]{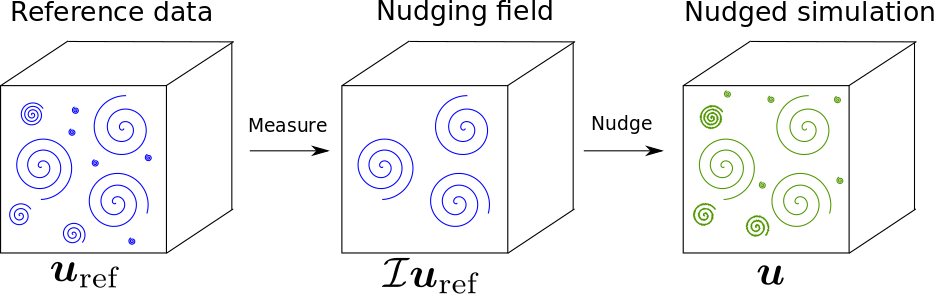}
    \caption{Diagram outlining the nudging algorithm. In our numerical
    experiments the reference data comes from a well controlled direct
    numerical simulation, and the process of measuring is summarized by
    the filtering operation $\mathcal{I}$.}
    \label{diagram}
\end{figure}

\begin{figure*}[t]
    \centering
    \includegraphics[width=0.8\textwidth]{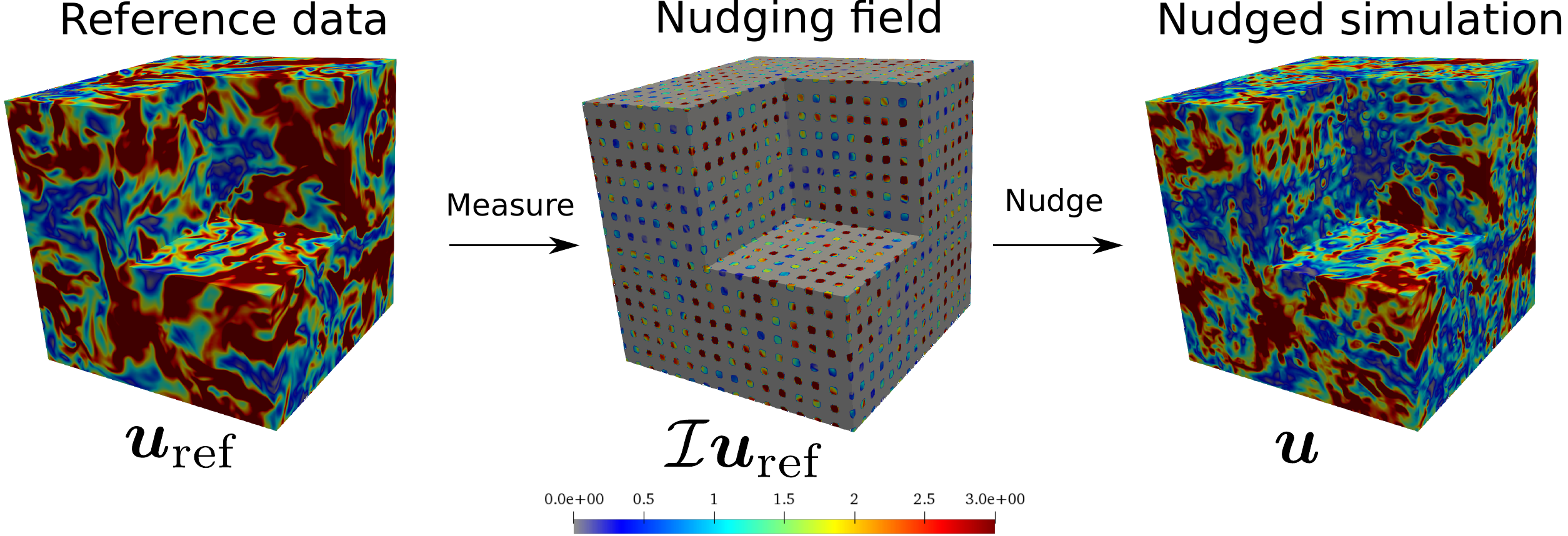}
    \includegraphics[width=0.8\textwidth]{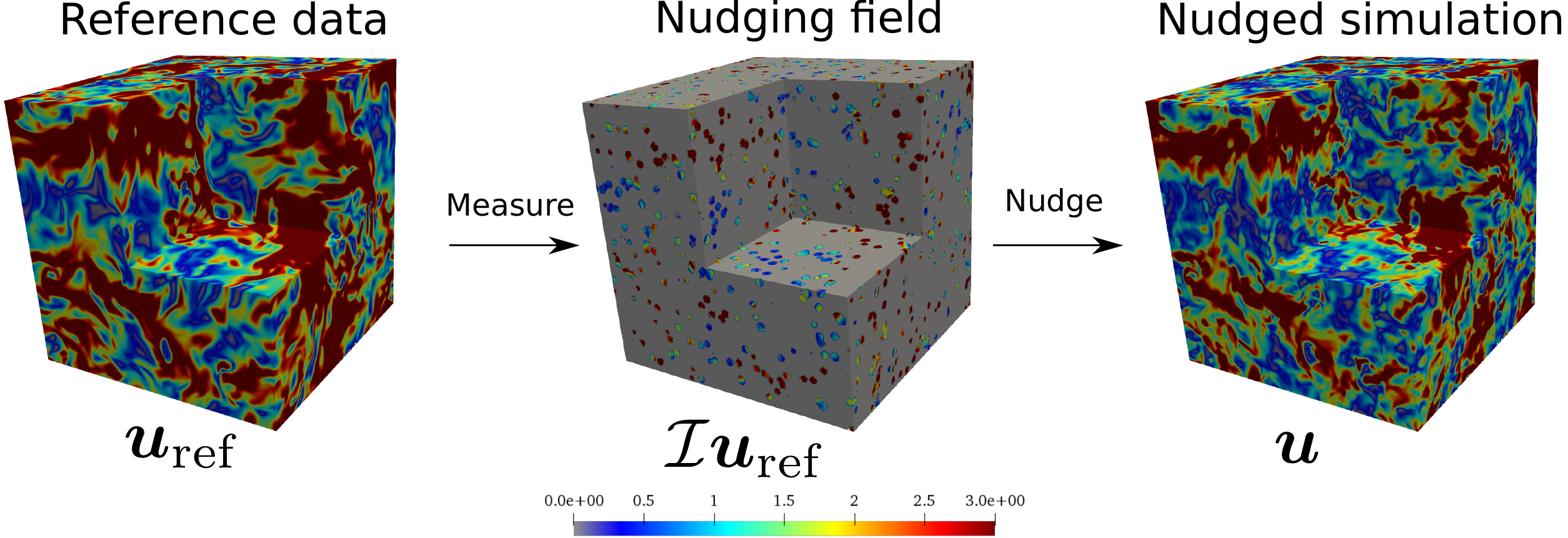}
    \caption{First row: visualizations of (left) reference fields
    $\uref$, (middle) filtered/nudging fields $\mathcal{I}\uref$, and (right)
    nudged/reconstructed fields $\bu$ for Eulerian nudging.  The parameters are
    $\alpha t_\eta=0.40$, $\tau/t_\eta=25$, $\phi=0.05$.  Second row the
    same of above but for Lagrangian nudging. The parameters are $\alpha
    t_\eta=0.40$, $\tau/t_\eta=25$, $\phi=0.03$.}
    \label{viz}
\end{figure*}

Given the problem of trying to reconstruct the whole  flow configuration  out of some partial data,
one may ask two crucial questions.  The first one is about the {\it
quantity} of information that one needs to collect in order to achieve a
certain degree of reconstruction. The second one concerns how the {\it
quality}, or type, of information affects the level of reconstruction
that can be attained.   The two main tools used in DA are based on
either variational or ensemble-averaged approaches.  Variational
methods, best exemplified by the 4D-Var technique
\cite{Talagrand87,Park03,Rawlins07}, rely on minimizing the distance
between a simulated system's trajectory with the available data. In
order to do this, the statistics of the errors are assumed to be
Gaussian.  Ensemble approaches work by performing Kalman filtering
operations \cite{Evensen94,Evensen,Houtekamer16} on the probability
distributions of different realizations of the state to be
reconstructed. Similarly to the variational approaches, they also assume
the statistics to be Gaussian. Both techniques have proven useful in NWP
and have also been applied to mildly turbulent channel flows
\cite{Bewley04,Suzuki12,Suzuki17}, but they have never been put to these
in fully-developed turbulence, where the small-scale velocity statistics
is  intermittent with fat and  non-Gaussian tails, and the system is
strongly out of equilibrium. This constitutes a big hurdle to overcome
also for NWP, as new technological developments in computational and
measuring tools allow weather forecast centers to reach resolutions where
three-dimensional turbulent convection becomes important
\cite{Leutwyler16,Yano18}, signaling we are entering an era where
non-linear DA schemes have to be put to use. One possible scheme is
Particle Filtering \cite{Gordon93}, which works like the Kalman filter
based approaches but without employing the Gaussianity assumption. This
scheme has already been put to test in two-dimensional barotropic flows
\cite{Vanleeuwen13} and weather models \cite{Poterjoy17}, showing better
results than linear DA schemes, but still presenting non-trivial
obstacles when scaling to high-dimensional systems.

\begin{figure*}
    \centering
    \includegraphics[width=0.32\textwidth]{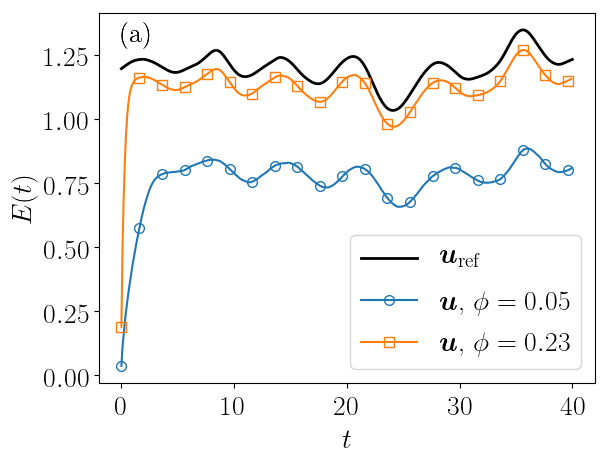}
    \includegraphics[width=0.32\textwidth]{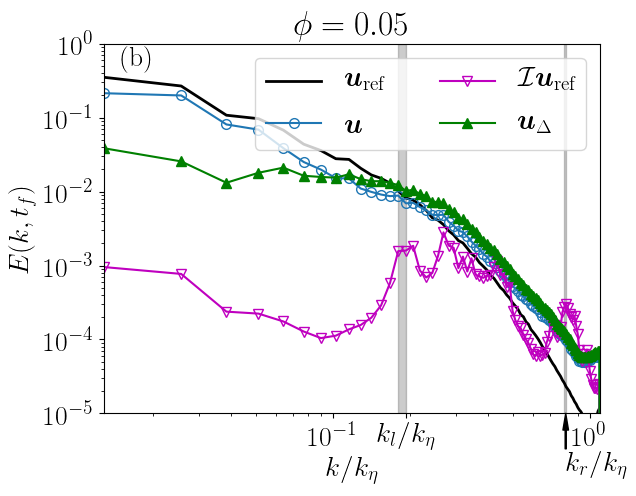}
    \includegraphics[width=0.32\textwidth]{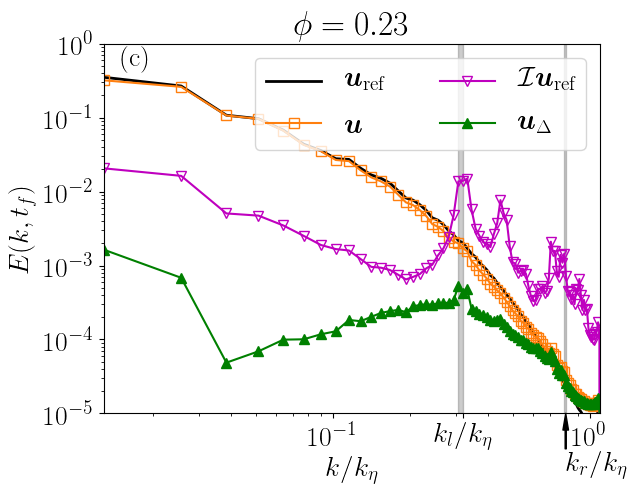}
    \caption{(a) Evolution of the total  energy for the reference field
    and for two nudged fields with different volume fraction, before and close to full synchronization, $\phi=0.05,0.23$.
    Log-log plots of (b) the energy spectra for the reference field (the {\it truth}), $\uref$), the nudging partial data, ${\cal I} \uref$ and the nudged/reconstructed field on the whole volume $\bu$. We also plot the spectrum of the error field, $\bu_\Delta$. 
    with $\phi=0.05$ and (c) for $\phi=0.23$. Grey regions mark the two
    typical wavenumbers  $k_l$ and $k_r$ (see text). Notice the transition to full
    synchronization for panel (c) where the spectrum of the error field,
    $\bu_\Delta$, is negligible at all scales.}
    \label{energy}
\end{figure*}

In this paper, we propose to use nudging
\cite{Hoke76,Lakshmivarahan13,Clark18}, a fully {\it unbiased} approach,
to numerically study the problem of assimilating data into a turbulent
flow which is characterized by a high (infinite) dimensional phase-space
with strong non-Gaussian and intermittent multi-scale fluctuations
\cite{Frisch,Pope}. Nudging has an old and prestigious past in DA
history \cite{Hoke76,Lakshmivarahan13}. It consists of applying a
penalty term to the right hand side of the evolution equations that
tries to minimize the distance between the evolved flow and the
observations (see Fig.~\ref{diagram} for a sketch). In a way, nudging
can be viewed as the application of a Newton relaxation feedback to
fluid flows. In the context of NWP, different formulations of nudging
have been used to study the state estimation problem using finite
dimensional dynamical systems and weather models
\cite{Hoke76,Auroux08,Du13,Pazo16}, and for boundary condition matching
\cite{Vonstorch00,Waldron96,Miguez-macho04}.  In the context of
turbulence, for the cases of two-dimensional Navier-Stokes Equation
\cite{Farhat16,Gesho16,Foias16,Biswas17}, the three-dimensional
Navier-Stokes $\alpha$ model \cite{Albanez16}, and Rayleigh-Bernard
convection \cite{Farhat17,Farhat19}, it has been rigorously proven that
given a sufficient amount of input data a nudged field will eventually
synchronize with its {\it nudging} field. Indeed,  both DA
\cite{Carrassi08} and nudging can be framed as a synchronization
problem, see  \cite{Lalescu13} for an application similar to Fourier
nudging  for turbulence.  

Before moving on, we should note that in the current data-driven age,
parameter reconstruction is another key problem for accurate flow
prediction, modelling, and control. Here the goal is to recover, out of
some given data, the form and/or the parameters of the underlying PDEs
(or ODEs) that generated such data. Modern methods include (but are not
restricted to) symbolic regression coupled with sparsity methods
\cite{Brunton16,Rudy17}, physics informed neural networks
\cite{Raissi17a,Raissi17b}, statistical inference \cite{Cialenco11}, and
minimum ignorance approaches \cite{Du12}.  Recently, we have shown that
nudging can be used to infer parameters and physics even for the case of
three dimensional fully developed turbulence, both isotropic and under
rotation \cite{Clark18}. Another related problem is the one of
equation-free modelling, where recent advances haven been made in high
dimensional systems by using reservoir computing techniques
\cite{Pathak18,Nakai18}, and in turbulent flows by using artificial
neural networks \cite{Mohan19}. All these problems point to the pressing
need of developing data-driven techniques that can be scaled to
non-linear and high-dimensional problems such as turbulence.

Our novel goal is to present nudging as a tool to probe for the key
degrees of freedom of a flow and understand {\it where} and {\it what}
we need to measure to ensure a certain level of reconstruction. We
tackle problems such as if it is better to (i) place the probes in a
regular equispaced way, (ii)  follow measurements in a Lagrangian
domain, along floating probes, or (iii) perform first a Fourier
convolution to spread the information on the whole configuration space. Furthermore, we will also study nudging in the presence of inverse energy cascade \cite{Alexakis18} with the formation of highly coherent cyclonic/anticyclonic structures in rotating turbulence.
These are the questions we will answer here, and that
have not been addressed before. Many others will follow, that we leave
for future research: What about bounded flows \cite{Kim87,Hwang16}? Is
it better to place the probes close to the wall or in the bulk? What
about multi-field equations as in Rayleigh-B\'enard
\cite{Fauve81,Ahlers09} or MHD \cite{Leorat81,Alexakis05}? Can we
control temperature by measuring velocity in convection? or velocity by
measuring the magnetic field in MHD? All these questions have applied
and fundamental importance.

The paper is organized as follows: in Sec.~\ref{methods} we outline the
how the nudging protocol works, in part \ref{nudging_methods} we write
down the equations, in \ref{numerical_protocols} we give details on the
numerical implementations of the technique and of the simulations
performed, and in \ref{error_def} we explain the different quantities we
will use to measure the performance of nudging. We will then present the
results of nudging in configuration space in \ref{conf_results}, of
nudging in Fourier space in \ref{four_results}, and of nudging under the
presence or large scale structures in \ref{large_scale_structures}.
Finally, we present our conclusions in \ref{conclusions}.

\section{Methods} 
\label{methods}

\subsection{Nudging the Navier-Stokes equations} 
\label{nudging_methods}

Our application of Nudging is based on the following protocol. Suppose
we have some measurements of a reference field data, $\uref$, available
only on certain regions of space (or for certain Fourier modes) and with
a certain cadence in time, $\tau$. And suppose we know that the field
evolution is described by the three dimensional incompressible
Navier-Stokes equations (NSE) with unit density:

\begin{equation}
\begin{cases}
    \partial_t {\bm u}_{\rm ref}  + {\bm u}_{\rm ref}
    \cdot {\bm \nabla} {\bm u}_{\rm ref} =  - {\bm \nabla} p_{\rm ref} +
    \nu \nabla^2 {\bm u}_{\rm ref} + \bm{f}_{\rm ref},     \label{nsref}\\
    {\bm \nabla} \cdot {\bm u}_{\rm ref}=0,\\
    + {\rm boundary \, conditions,}
    \end{cases}
\end{equation}
where $\bm{f}_{\rm ref}$ is a forcing mechanism and $\nu$ the
viscosity. The aim is to reconstruct the whole space-time evolution of
${\bm u}_{\rm ref}$ by evolving a numerical simulation for another
incompressible velocity field ${\bm u}$, which we call the nudged field,
where the distance from the input data ${\bm u}_{\rm ref} - {\bm u}$
enters as a penalty term: 

\begin{equation}
\begin{cases}
    \partial_t \bu  + \bu \cdot \bnabla
    {\bm u} =  - \bnabla p + \nu \nabla^2 \bu  - \alpha
    \mathcal{I}  ( \bu - \bu_\mathrm{ref}),     \label{nsnudged}\\
     {\bm \nabla} \cdot {\bm u}=0\\
    + {\rm boundary \, conditions,}
    \end{cases}
\end{equation}
where $\alpha$ is the amplitude of the nudging term, and $\mathcal{I}$
is a filtering operator which projects $\bm{u}-\bm{u}_{\rm ref}$ onto
the regions of space (or the Fourier scales) in which the reference data
is known. We refer to $\mathcal{I} \uref$ as the nudging field.  If the
cadence, $\tau$,  at which the observations are available does not
coincide with the time step used to evolve the nudged system, one then
has to define a reference field $\uref^\tau$ time interpolated between
the two consecutive measurements. There are two very important aspects
to be noted here. First, nudging can in principle be formulated for any
dynamical system or PDE, as for example done by
\cite{Azouani13,Azouani14}, i.e. its formulation does not depend on the
application to the NSE.  Second, the term $\bm{f}_{\rm ref}$ can be
quite general, it does not have to be just a simple mechanical injection
mechanism, it could also depend on $\uref$ for example.  The filter
operator $\mathcal{I}$ can take many forms too.  The first one that we
address here is based on local measurements of the velocity field:

\begin{equation}
    \mathcal{I} \bm{u}(\bm{x},t) = \sum_{i=1}^{N_p} \int
    \bm{u}(\bm{x},t) \delta({\bm x} - {\bm X}_i(t)) dV,
    \label{conf_filt}
\end{equation}
where $\bm{X}_i$(t) are the positions of the $N_p$ probes where the
input data are measured, and that can be fixed in space (Eulerian case)
or moving with the flow (Lagrangian case). Our implementation of
(\ref{conf_filt})  will actually act on small volumes and will be
referred to as ``configuration space nudging'' (more on this in
Sec.~\ref{numerical_protocols}). The second family of nudging protocols
that we study here is based on  a Fourier filtering:

\begin{equation}
    {\cal I} {\bm u}(\bm{x},t) = \sum_{ {\bm k} \in {\cal A}}
    \hat{{\bm u}}({\bm k},t) \exp{(i {\bm k}\cdot {\bm x})},
    \label{four_filt}
\end{equation}
where $\hat{\bm{u}}({\bm k})$ are the Fourier coefficients of the field
$\bu$, and ${\cal A}$ is a given sub-set of the Fourier space where we
suppose to know the evolution of the reference field coefficients,
$\hat{{\bm u}}_{\mathrm{ref}}({\bm k})$.  While in principle the set
${\cal A}$ can be arbitrary, in this work we  will always use  a
low-pass filter: 

\begin{equation} 
    {\cal A}: \{ |{\bm k}| < k_n \}, 
    \label{eq:low}
\end{equation}
i.e.  we will nudge a band  of large-scale modes in the flow.
Simulations performed using this filter will be referred to as
``spectral nudging''.  It is very important to notice that we are
playing the reconstruction game in a {\it fair} way, without assuming to
know anything about the external forcing mechanisms that has generated
the reference field in \eqref{nsref}. This is the minimal set-up if we
want to be realistic (in most applications even the boundary conditions
are not fully under control and certainly not the space-time
configuration of the external stirring force).   This set-up  will
prevent us from reaching any exact synchronization of the two fields
because $\bu_\mathrm{ref}=\bu$ is  not a solution of \eqref{nsnudged}
anymore, but it  allow to speak about a real-life problem. The absence
of a forcing stirring term in \eqref{nsnudged} also implies that without
nudging the reconstructed flow would decay to zero monotonically, as we
inject energy only by the information coming from the $  \mathcal{I}  (
\bu - \bu_\mathrm{ref}) $ term.

\begin{figure}
    \centering
    \includegraphics[width=8.5cm]{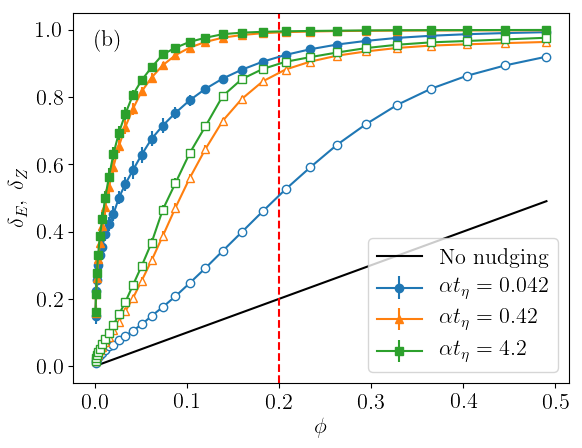}
    \includegraphics[width=8.5cm]{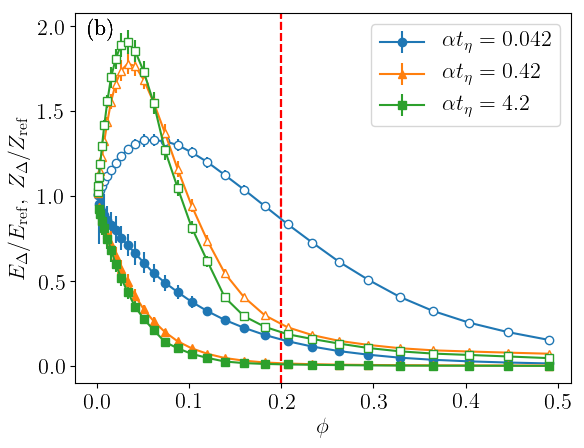}
    \caption{(a): Values of the velocity correlations $\delta_E$ (filled
    symbols) and the vorticity correlations $\delta_Z$ (empty symbols).
    (b) Values of the energy error $E_\Delta/E_{\rm ref}$ (filled
    symbols) and the enstrophy errors $Z_\Delta/Z_{\rm ref}$ (empty
    symbols). All values are plotted at changing volume fraction $\phi$
    for different values of the nudging amplitude $\alpha$. Solid line
    in panel (a) represents the linear scaling expected for no effects
    of nudging (see text). The dashed red vertical lines mark the value
    $\phi_c=0.2$ as the typical estimate for transition to maximum
    synchronization in these sets-up. Here and in all figures error bars
    are always plotted, when they are not visible it means that they are
    smaller than the symbol size.}
    \label{scan_real}
\end{figure}

\begin{figure}
    \centering
    \includegraphics[width=8.5cm]{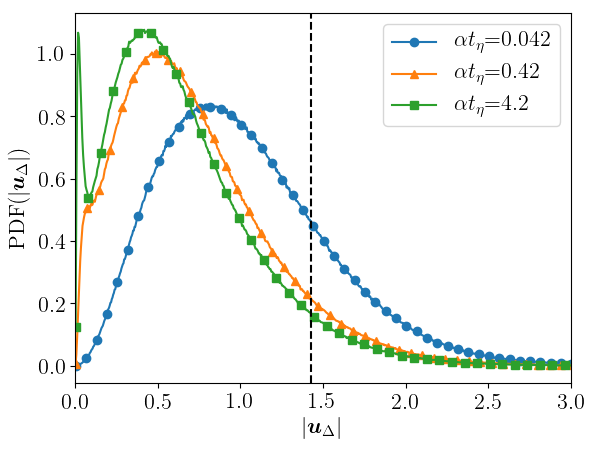}
    \includegraphics[width=8.5cm]{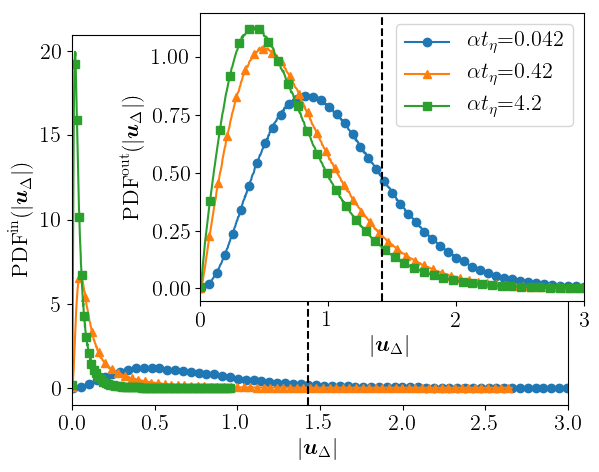}
    \caption{Histograms of the point-by-point reconstructing error
    $|\bm{u}_\Delta|$ measured in the whole volume (top), only inside
    the nudging regions (bottom) and only outside (inset). The vertical
    dashed line represent  $\langle |\uref| \rangle$.}
    \label{histograms}
\end{figure}

\begin{figure}
    \centering
    \includegraphics[width=8.5cm]{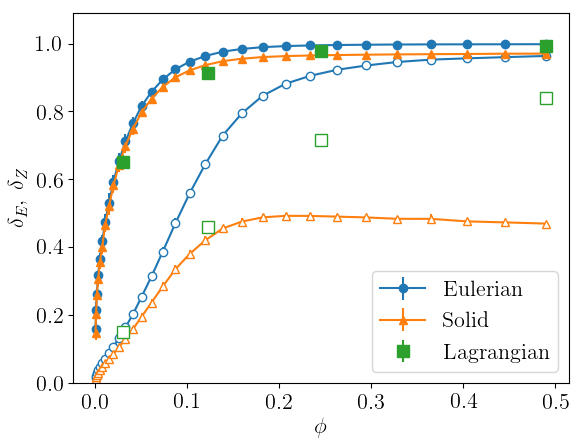}
    \caption{Values of the velocity correlations $\delta_E$ (filled
    symbols), and the vorticity correlations $\delta_Z$ (empty symbols)
    as a function of the volume fraction $\phi$ for different 
    nudging protocols.}
    \label{scan_solid}
\end{figure}

\begin{figure*}[t]
    \centering
    \includegraphics[width=0.32\textwidth]{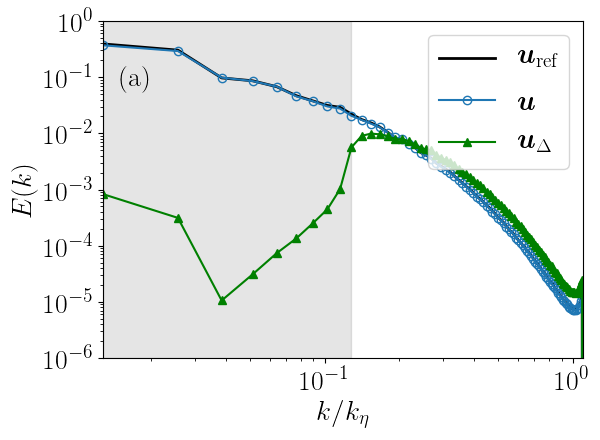}
    \includegraphics[width=0.32\textwidth]{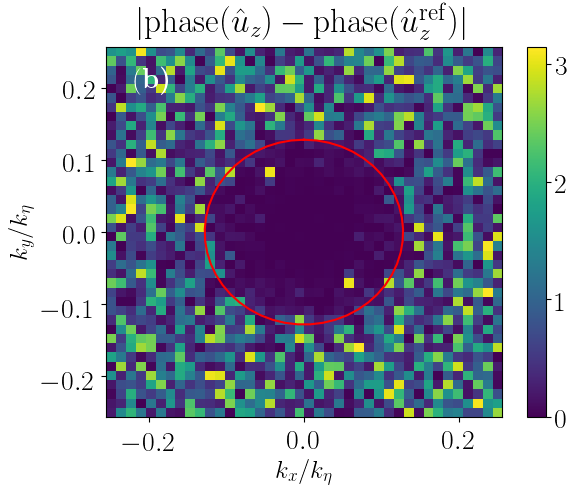}
    \includegraphics[width=0.32\textwidth]{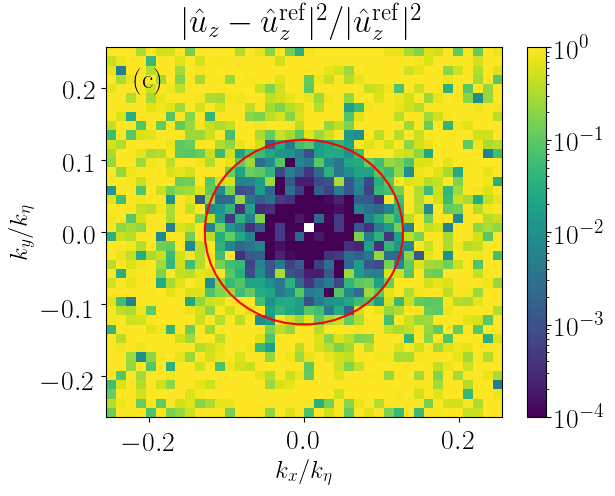}
    \caption{(a) Energy spectra for the nudged/reconstructed, $\bu$,
    the reference, $\uref$ and the error, $\bu_\Delta$ fields.  The grey
    area indicates the nudged scales, $ k<k_n$. (b) Point-by-point error
    between the phases of the nudged/reconstructed field and the
    reference one in the $k_z=0$ Fourier plane. (c) The same of panel
    (b) but for the normalized amplitudes of the z component. The red
    circle defines $k_n$}
    \label{spectral}
\end{figure*}

\subsection{Numerical protocols} 
\label{numerical_protocols}

\begin{table}
    \centering
    \begin{tabular}{c | c | c | c | c | c | c | c }
        Type      & $E_{\rm ref}$ & $\mathrm{Re}$ & $\nu$       &
        $t_L$     & $t_\eta$      & ${k}_{\eta}$  & $N^3$
        \\
        \colrule
        RUN1  & 1.20          & 3900          & 0.0025      &
        4.06      & 0.0042        & 78            & $256^3$
        \\
        RUN2      & 1.27          & 25000         & 0.0004      &
        3.94      & 0.00062       & 317           & $1024^3$
    \end{tabular}
    \caption{Parameters used for the different reference simulations
    experiments. All the respective nudged simulations had the same
    parameters. The code uses a two step Adams Bashfort scheme for the
    time integration, and the ``2/3 rule'' for dealiasing. The values
    listed are the total kinetic energy $E=1/2 \langle |\uref|
    \rangle^2$, the Reynolds number $\mathrm{Re}= L (2 E)^{1/2}/\nu$,
    the viscosity $\nu$, the eddy turnover time $t_L=L/(2 E)^{1/2}$, the
    Kolmogorov timescale $t_\eta = \nu L/(2 E)^{3/2}$, the Kolmogorov
    wavenumber $k_\eta=(\nu^3 L/(2 E)^{3/2})^{-1/4}$, 
    and the number of grid points $N^3$. The largest scale of the flow
    $L$ is equal to $2\pi$ in all simulations. In both cases
    $\bm{f}_{\rm ref}$ is a randomly-generated, quenched in time,
    isotropic field with support on wavenumbers with amplitudes $
    k\in[1,2]$ whose Fourier coefficients are given by
    $\bm{\hat{f}}_{{\rm ref}}({\bk}) = f_0 k^{-7/2} e^{i\theta_{\bm
    k}}$, where $\theta_{\bm k}$ are random in $[0,2\pi)$ and $f_0=0.02$.}
    \label{table1}
\end{table}

In our  study, the reference {\it true} data $\bm{u}_{\rm ref}$ is
generated by numerically solving the Navier-Stokes equations
\eqref{nsref}, instead of using experimental measurements or field
observations. The obvious advantage is that we can benchmark the
reconstruction capabilities of nudging in a fully quantitative way, as
we have access to the truth in every point in space and at every scale.
Two different reference sets were produced, at medium and high Reynolds
number (see Table~\ref{table1} where all the details of the numerical
methods used to solve \eqref{nsref} and \eqref{nsnudged} are given). In
the rest of the paper, all values are made dimensionless by fixing the
kinetic energy, the size of the box, and the viscosity.  The exact
protocol adopted is the following. Starting from rest, we evolve
\eqref{nsref} until the system reaches a stationary state (marking this
moment as $t=0$). Then we run for 10 turn over times (marking the final
moment $t=T$), saving the fields at high frequency. We then solve
\eqref{nsnudged} in the interval $t\in[0,T]$, using as initial condition
$\mathcal{I} \bm{u}_{\rm ref} (\bm{x},t=0)$ and inputting the linearly
interpolated field $\bm{u}^\tau_{\rm ref}$ into the nudging term. This
is done for different values of $\alpha$ and $\tau$, and for the
different filters $\mathcal{I}$ (Configuration Eulerian/Lagrangian or
Fourier).  The implementation of the point measurement based filter,
\eqref{conf_filt}, is a bit delicate. As we do not have any other
injection mechanism in \eqref{nsnudged}, nudging only in points (i.e.
one grid point) makes it difficult to inject enough energy in order to
maintain a stationary simulation with comparable $Re$. For this reason
we actually nudge in small spheres of radius $r=1.25\eta$ centered
around points $\bm{X}_i$. For the Eulerian set-up, these points were
always placed on a uniform equispaced three dimensional grid covering
the whole simulation box, so the only controlling parameter is the total
number of probes $N_p$. The number we use to characterize each grid is
the volume fraction:

\begin{equation}
    \phi = N_p \frac{(4/3) \pi r^3}{L^3},
\end{equation}
which is the ratio between the nudged and the total volumes. There are
two useful wavenumbers that can be defined:

\begin{equation}
    k_l = \frac{2\pi N^{1/3}_p}{L},\qquad    k_r = \frac{2\pi}{r}.
\end{equation}
where $k_l$  is associated with the minimum distance between probes and
$k_r$ with the probe size.  For the Lagrangian set-up, the protocol is
similar with the only difference that the probe positions will move in
time following the equation of a fluid tracer: 

\begin{equation}
    \bm \dot X_i(t) = \bu_{{\rm ref}}({\bm X}_i(t),t) .
\end{equation}

In Fig.~\ref{viz}  we give a first qualitative anticipation of both
protocols, showing a 3D rendering of the reference field, of the probe
distributions (nudging stations) and of the reconstructed flow for both
Eulerian and Lagrangian nudging at high Reynolds.  \noindent As a third
variation, we will also explore nudging with spherical probes (placed on
an Eulerian grid) where the velocity is fixed to have the same value of
the one assumed in the center, making the filtered field $\mathcal{I}
\uref$ piece-wise constant and mimicking the results from a localized
reference field measurement. We refer to this scheme as ``solid''
nudging.

\subsection{Quantification of errors and correlations} 
\label{error_def}

We start by defining the difference between the two fields (error field)
at every space-time point:

\begin{equation}
    \bm{u}_\Delta (\bm{x},t) = \bm{u}(\bm{x},t) - \uref(\bm{x},t).
\end{equation}
Then, in order to quantify the nudging performances for turbulent DA at
both large and small-scales we define the relative errors in the
point-to-point energy and enstrophy reconstruction, based on the
time-averaged  $L_2$ norm: 
\begin{equation}
\label{eq:E}
    \frac{E_\Delta}{E_{{\rm ref}}} =
    \frac{\langle|\bm{u}_\Delta|^2\rangle}{\langle|\uref|^2\rangle},
    \qquad
    \frac{Z_\Delta}{Z_{\rm ref}} = \frac{\langle|{\bm
    \omega}_\Delta|^2\rangle}{\langle|{\bm \omega}_{\rm
    ref}|^2\rangle},
\end{equation}
where ${\bm \omega}=\bm{\nabla}\times\bu$ is the  vorticity field and
the average is defined as the mean on the whole volume, $V$, and on the
whole experiment duration, $T$:  $ \langle \bullet \rangle = 1/(T \,V)
\int_0^T dt \int_V d\bx \, (\bullet)$. We sometimes look at the temporal
variations too, in those cases we explicitly remark that what we are
showing depends on time. So, for example, the time evolution of the
energy of a nudge simulation will be referred to as $E(t)$.

In order to have a scale-by-scale control of the degree of
synchronization we introduce  the energy spectrum of the difference
between the nudged and the reference field, given by

\begin{equation}
    E_\Delta(k,t) = \frac12 \sum_{k \le |\bk|<k+1}
    |\hat{\bm{u}}_\Delta(\bk,t)|^2.
    \label{edelta}
\end{equation}

The two most informative measures of the success of reconstruction at
large/small scales are based on velocity/vorticity field correlations

\begin{equation}
\label{eq:delta}
    \delta_E = \frac{ \langle \bm{u} \cdot \bm{u}_{\rm ref} \rangle }{
    \langle |\bu| \rangle \langle |\bu_{\rm ref}| \rangle },\qquad
    \delta_Z = \frac{ \langle \bm{\omega} \cdot \bm{\omega}_{\rm ref}
    \rangle }{\langle|{\bm \omega}|\rangle \langle|{\bm \omega}_{\rm
    ref}|\rangle}.
\end{equation}
Both quantities,  will give a good account of how much the nudged fields
is close to the reference one independently of the absolute values of
each field (see later). Evidently, we have:

\begin{equation}
    -1 \le \delta_E, \delta_Z \le 1.
\end{equation}

Finally, it will be instructive to look also at the probability
distribution function (PDF) of the point-wise error, $|{\bm
u}_\Delta(\bx,t)|$ in order to understand specific issues connected to
{\it worst-case} scenarios and/or whether there are spatial and/or
topological structures  that are better reconstructed. The latter point
might not be so relevant for  isotropic turbulence but it is a key issue
in non-isotropic conditions, like in the presence of boundaries or large
scale shear, as often happens in nature or in applied turbulent
realizations.  

\begin{figure}
    \centering
    \includegraphics[width=8.5cm]{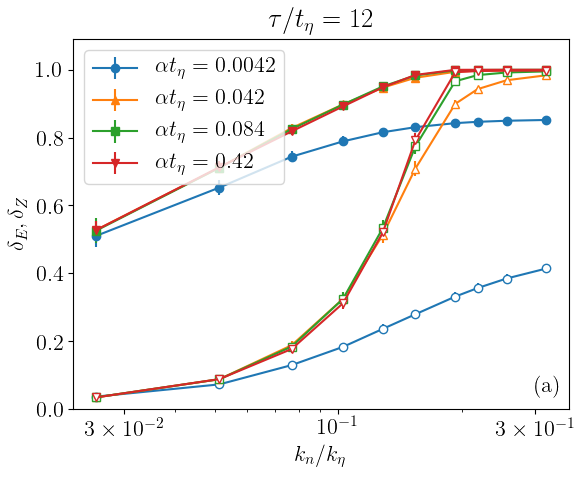}
    \includegraphics[width=8.5cm]{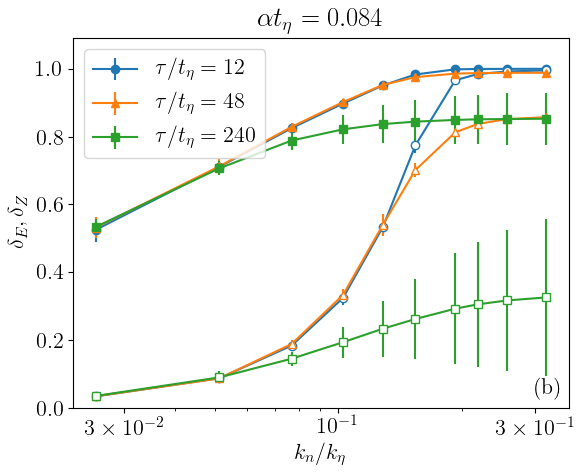}
    \caption{Values of the velocity correlations $\delta_E$ (filled
    symbols), and the vorticity correlations $\delta_Z$ (empty symbols)
    as a function of the maximum nudged wavenumber, $k_n$, for different
    values of the nudging amplitude, $\alpha$, and the interpolation
    time, $\tau$. In (a)  $\tau/t_\eta =12$, in (b) $\alpha
    t_\eta=0.084$.}
    \label{scan_spectral}
\end{figure}

\begin{figure}
    \centering
    \includegraphics[width=8.5cm]{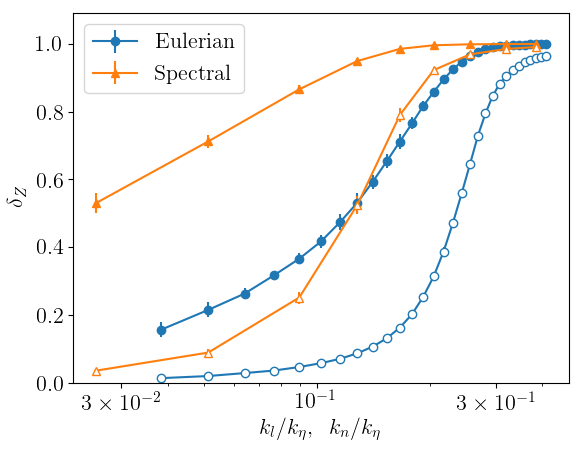}
    \caption{Values of the velocity correlations, $\delta_E$ (filled
    symbols), and the vorticity correlations, $\delta_Z$ (empty
    symbols), as a function of the maximum nudged wavenumber, $k_n$
    (spectral case), or the mean nudged wavelength, $k_l$ (Eulerian
    case).}
    \label{real_vs_spectral}
\end{figure}

\section{Results} 
\label{results}

\subsection{Nudging in configuration space} 
\label{conf_results}

We start by studying the case of nudging in configuration space, where
the penalty term acts in confined regions in space.  From Fig.~\ref{viz}
we qualitatively see that the nudged flows (right panels) develop
large-scale structures very close to the reference fields (left), even
though nudging only acts locally. In this section we will focus on the
effects of varying the nudging amplitude $\alpha$ and the nudged volume
fraction $\phi$. In all simulations, the temporal interpolation
$\tau/t_\eta = 25$ and only data from RUN1 were used (see
Table~\ref{table1}).  We will study the response at varying the
time-interpolation cadence, $\tau$, in Sec.~\ref{four_results}.

In Fig.~\ref{energy}a we show the evolution of the total energy for two
nudged simulations,  with $\phi = 0.05$ and  $\phi = 0.23$. Both have
$\alpha t_\eta=0.42$. The evolution of the total reference energy is
also shown. As explained above, the initial condition of the nudged
simulations is given by filtered reference at $t=0$, so they would look
just like the middle panel in Fig.~\ref{viz}.  It takes about one eddy
turn over time for the nudged simulations to reach the stationary state,
and to synchronize with the reference evolution, as seen in
Fig.~\ref{energy}a. The evolution of the energy shows some very
interesting features. First, the energy of the nudged field is always
smaller than that of the reference field. Second, nudging a higher
volume fraction does indeed inject more energy and make the nudged
system resemble the reference one more closely.  It is important to
remember that aside from the nudging term, no energy is being injected
in the simulations as there is no external forcing mechanism present in
\eqref{nsnudged}. Third --and probably most striking--, the nudged
simulations is always able to follow the dynamical fluctuations of the
reference field even in the presence of an appreciable amplitude
mismatch. The latter, is the indication that we can have good
statistical correlations among the two fields without complete
synchronization. This will be put in more quantitative terms below.

\begin{figure}
    \centering
    \includegraphics[width=8.5cm]{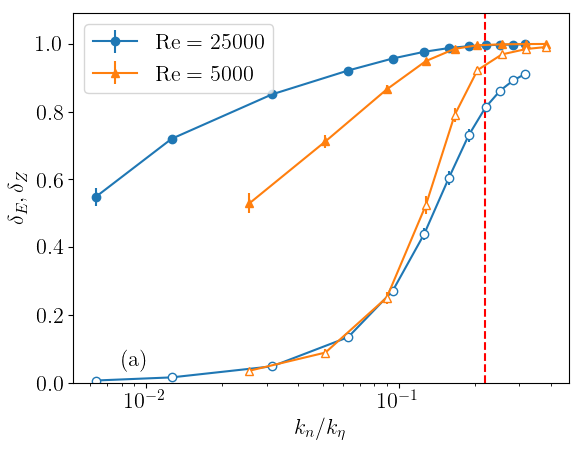}
    \includegraphics[width=8.5cm]{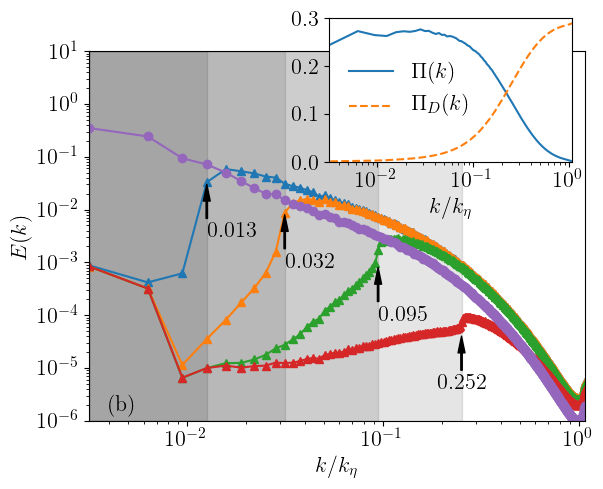}
    \caption{(a) Values of the velocity correlations $\delta_E$ (filled
    symbols), and the vorticity correlations $\delta_Z$ (empty symbols)
    as a function of the maximum nudged wavenumber $k_n$ for different
    values of the nudging amplitude $\alpha$ and the interpolation time
    $\tau$. Red dashed line marks the $k_c \sim 0.2 k_\eta$ value where
    transition-to-synchronization is obtained  (b) Energy spectra of the
    reference simulation RUN2 (round markers) and spectra of the
    difference field $u_\Delta$ at different $k_n$ (triangular markers).
    The value of $k_n/k_\eta$ for each nudged simulation is annotated with
    arrows.  Inset: non-linear and viscous contributions to the  energy
    fluxes for the reference  simulation RUN2.}
    \label{re_and_flux}
\end{figure}

\begin{figure*}
    \centering
    \includegraphics[width=0.9\textwidth]{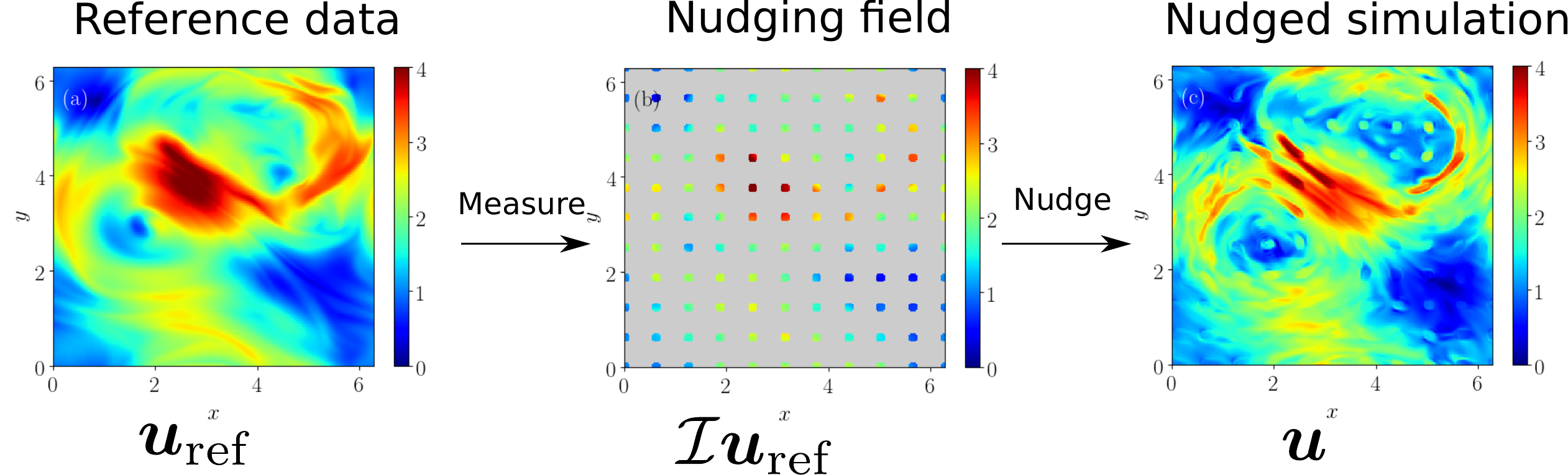}
    \caption{Visualizations of the local  energy field of (a) a
    reference simulation $\uref$, (b) a filtered field
    $\mathcal{I}\uref$, (c) and a nudged simulation $\bu$ of a rotating
    turbulent flow for $\phi=0.15$, $\alpha t_\eta = 0.3$, and
    $\tau/t_\eta=33$. The reference simulation has $E_{\rm ref} = 1.3$,
    $\nu=0.002$, $N^3=256^3$, $L=2\pi$, $t_L = 3.89$, $t_\eta  = 0.003$,
    $\mathrm{Re}=5000$, and $\mathrm{Ro} = 0.06$. The forcing,
    $\bm{f}_{\rm ref}$, is a randomly-generated, quenched in time,
    isotropic field with support on wavenumbers with amplitudes $
    k\in[1,2]$ whose Fourier coefficients are given by
    $\bm{\hat{f}}_{{\rm ref}}({\bk}) = f_0 k^{-7/2} e^{i\theta_{\bm
    k}}$, where $\theta_{\bm k}$ are random in $[0,2\pi)$ and
    $f_0=0.005$.}
    \label{rot_viz}
\end{figure*}

In Figs.~\ref{energy}b and \ref{energy}c we compare the instantaneous
energy spectra of (i) the total reference field, $\bu_\mathrm{ref}$;
(ii) the filtered reference field used for nudging, $ \mathcal{I}
\bu_\mathrm{ref}$;  (iii) the  resulting nudged field, $ \bu$ and the
one which quantify the synchronization error (\ref{edelta})  for two
different nudging volume fractions $\phi=0.05$ and $\phi=0.23$,
respectively. First, let us notice that the spectrum of $ \mathcal{I}
\bu_\mathrm{ref}$ is mainly concentrated at small scales (large
wavenumbers), with peaks in correspondence of the minimum distance
between probes, $k_l$, and of the probe size, $k_{r}$, indicating that
we are not supplying a large amount of information concerning the global
large-scale motion (small wavenumbers). Despite of this,  the
scale-by-scale synchronization error, $E_\Delta(k,t)$, is smaller at
large scales (small wavenumbers) than at  small scales (large
wavenumbers).  Furthermore, in the case with $\phi = 0.23$ (panel c) the
errors remain small across all scales, indicating a very good global
reconstruction and a transition to {\it full  synchronization}  already
for such relatively small volume fraction. 

In Fig.~\ref{scan_real} we show, for 3 different  values of $\alpha$,
the correlations $\delta_E$ and $\delta_Z$, given by (\ref{eq:delta})
and the normalized errors $E_\Delta/E_{\rm ref}, Z_\Delta/Z_{\rm ref}$
given by (\ref{eq:E}) as a function of $\phi$. Good correlations and
small relative errors in both the velocity and vorticity fields can be
achieved with  small nudged volume fractions. As one can see from the
panel (a), already at $\phi \sim 0.2$ and for a nudging coefficient
strong enough ($\alpha t_\eta \sim 0.5$)  we  can reconstruct both total
energy and total enstrophy with an accuracy close to $90\%$.  As
expected, $\delta_E$ converges faster than $\delta_Z$, as it is
determined by the large scales. For very small volume fractions, $\phi
\ll 0.1$, the error is large, $E_\Delta/E_{\rm ref}=Z_\Delta/Z_{\rm
ref}=1$, as the nudging field is almost equal to zero due to the fact
that very little energy is injected into the system. As more energy is
injected, $\phi \sim 0.05$  the relative error in the enstrophy increase
at the beginning while the one in the energy always decreases. This is
because the velocity field can generate correlations more easily, while
the vorticity field does not, so for $\phi \ll 0.1 $ one gets:

\begin{align*}
    \frac{Z_\Delta}{Z_{\rm ref}} &= \frac{\langle|{\bm \omega} -
    {\bm \omega}_{\rm ref}|^2\rangle}{\langle|{\bm \omega}_{\rm ref}|^2\rangle}
    \approx
    \frac{\langle|{\bm \omega}|^2\rangle + | \langle
    {\bm \omega}_{\rm ref}|^2\rangle}{\langle|{\bm \omega}_{\rm ref}|^2\rangle}
    \approx 2,
\end{align*}
where we have used that $\langle {\bm \omega} \cdot {\bm \omega}_{\rm
ref} \rangle \approx 0$ and that $\langle|{\bm \omega}|^2\rangle \approx
\langle|{\bm \omega}_{\rm ref}|^2\rangle$.  By comparing the behaviour
for the three different values of $t_\eta \alpha$  one sees that by
increasing $\alpha$ the transition to synchronization becomes sharper
and little improvement is obtained as soon as $\alpha$ is of the same
order of the highest frequency in the turbulent flow, $ \sim 1/t_\eta$.
The key parameter that drives the transition to synchronization is the
volume fraction, and we can estimate the saturation to maximum
achievable reconstruction for $$\phi_c \sim 0.2$$.  In
Fig.~\ref{scan_real} we also plot the {\it naive} expectation obtained
by supposing that nudging works only where we supply the information and
gives fully uncorrelated results otherwise. In this case, the
correlation coefficients would just scale as the volume fraction, $\phi$
(solid line in Fig.~\ref{scan_real}a).

In Fig.~\ref{histograms} we show the PDF for the point-by-point error,
$|\bu_\Delta|= |\bu-\uref|$, with the statistics taken over the whole
volume (top panel),  only inside the nudged regions (bottom panel), and
only outside the nudged regions (inset), for different values of $\alpha
t_\eta$ and with  $\phi=0.05$. In accordance with Fig.~\ref{viz}, the
errors inside the nudged regions are usually quite small, especially if
compared with the mean $\langle |\uref| \rangle$, denoted by the dashed
vertical line in each figure.  The statistics outside the nudged regions
dominate the statistics over the whole volume, as the volume fraction is
small. Increasing $\alpha$ pushes the mean and the mode of the errors
closer to zero, without producing any fat tails in the distribution.

Finally, in Fig.~\ref{scan_solid} we compare the three different ways of
performing nudging in configuration space described in
Sec.~\ref{nudging_methods}, Eulerian nudging, Solid nudging, and
Lagrangian nudging, by looking $\delta_E$ and $\delta_Z$ as function of
the nudging volume fraction.  In all three cases $\alpha t_\eta=0.40$
and $\tau/t_\eta=25$. As one can see from $\delta_E$, the velocity field
gets well reconstructed by all  schemes. On the other hand, $\delta_Z$
indicates that vorticity reconstruction does not work well for the
"solid" schemes, as one could have expected because of the lack of
small-scales information for this case.  Surprisingly, also Lagrangian
nudging performs slightly worse. One possible explanation is that the
movement of the probes does not leave enough time for the flow
synchronization  at each point.  One possible way to fix this problem
could be to implement delayed-coordinates nudging, where the past
history of the data is also used at each instant to guide the
reconstruction, as was proposed for much simpler dynamical systems in
\cite{Pazo16} and never applied to turbulence up to now. 

\subsection{Nudging in Fourier space} 
\label{four_results}

We now turn to characterize how nudging in Fourier space works. We
analyze the effects of varying the nudging amplitude $\alpha$, the
interpolation time $\tau$ and the maximum nudged wavenumber $k_n$ in
(\ref{eq:low}). To get a first glimpse of spectral nudging, we show in
Fig.~\ref{spectral}a the  instantaneous energy spectrum for the full
reference field, $\uref$, that of the corresponding
nudged/reconstructed field, $\bu$, and the scale-by-scale
synchronization error, $E_\Delta(k,t)$, for a simulation with $\alpha
t_\eta=0.042$, $\tau/t_\eta=25$, and $k_n/k_\eta=0.13$. The grey region
indicate the nudged window $k \in [0:k_n]$. Nudging is able to
synchronize the nudged scales correctly as seen by the fact that
$E_\Delta(k,t)$ is very small for $k < k_n $, and also in
Figs.~\ref{spectral}b and ~\ref{spectral}c, where the synchronization
error for an instantaneous realization of  Fourier phases and amplitudes
are shown, respectively.  The red circle in Figs.~\ref{spectral}b and
~\ref{spectral}c denotes the maximum nudged wavenumber $k_n$.
Concerning the {\it transition to synchronization} we study now what
happens at changing $k_n$.  Figure~\ref{scan_spectral} shows the
equivalent of Fig.~\ref{scan_real} but for Fourier nudging, i.e.
$\delta_E$ and $\delta_Z$, as a function of $k_n/k_\eta$ for different
values of $\alpha$ while keeping $\tau$ fixed (panel a), and for
different values of $\tau$ while keeping $\alpha$ fixed (panel b).
Velocity field correlations start at high values, already for small
$k_n$, as the smallest wavenumbers contain most of the energy, but
vorticity field correlations require a larger amount of modes to be
nudged in order to build up. At around $$k_n = k_c \approx 0.2 k_\eta,$$
both $\delta_E$ and $\delta_Z$ show perfect synchronization being
both equal to one.

By looking at Fig.~\ref{spectral}a, one recognize that $k/k_\eta=0.2$
is around the end of the inertial range, indicating that one has to
nudge everything but the viscous modes in order to reach the
transition-to-synchronization limit. A similar result was found by
\cite{Lalescu13}, where, at difference from here, synchronization was
studied by imposing  the nudged modes  to be equal to the reference ones
(something similar to $\alpha \to \infty$) and by supplying also the
exact external forcing field. It is important to remark that  the
$\#_{dof}$ necessary to  control for full synchronization, $k_c \sim 0.2
k_\eta$ implies that the number of modes being nudged is still much
smaller compared to the total number of {\it dof}, around $1\%$
actually, as the system is three dimensional.  From
Fig.~\ref{scan_spectral} we also see that similar to the case of nudging
in configuration space, increasing $\alpha$ has a positive effect. As
expected, decreasing $\tau$ has a negative effect.  The smaller the
scales that we nudge, the more sensitive they become to the choice of
$\tau$. This is because each Fourier mode has a characteristic
correlation time, given the sweeping time $\tau_s(k) \sim 1/(\sqrt{2E}
k)$ \cite{Chen89,Clark14,Clark15}, that becomes shorter the higher the
wavenumber. So if the correlation time of a particular mode becomes
shorter than the interpolation time $\tau$, the interpolation starts to
introduce unwanted errors. In Figure~\ref{real_vs_spectral} we show the
value of $\delta_E$ and $\delta_Z$ for Fourier  nudging as a function of
$k_n/k_\eta$ and for configurations space nudging as a function of
$k_l/k_\eta$. The functional behaviour is very similar, indicating that
both Fourier and configuration degrees of freedoms play a similar role
in driving the chaotic evolution of isotropic turbulence.  In other
words, there are not preferred {\it leading} variables that drive the
global and local flow configuration. The situation can be obviously very
different, whenever the flow is driven by boundary effects as in channel
turbulence, external fields, as for convection and MHD or influenced by
the global set-up as for rotation (see
Sec.~\ref{large_scale_structures}).

The effect of increasing the Reynolds number is studied in
Fig.~\ref{re_and_flux}a, where we compare velocity and vorticity
correlations for RUN1 and RUN2 (see Table~\ref{table1}) as a function of
$k_n/k_\eta$.  The fact that these two scans collapse on top of each
other when plotting against $k_n/k_\eta$ shows that $k_\eta$ is the
determining scale here. This can be understood better when looking at
the energy spectra and flux. Figure~\ref{re_and_flux}b shows the energy
spectra when nudging at different $k_n$ for the high Reynolds case. We
see that when correlations are high, the spectra of the differences
stays small for non-nudged wavenumbers. The inset of the figure shows
the non-linear, $\Pi(k)$, and dissipative, $\Pi_D(k)$ contributions to
the energy flux \cite{Alexakis18} for the reference simulation (RUN2).
The value of $k_n/k_\eta$ for which synchronization is achieved is the
same value at which the dissipation flux and the energy flux become
equivalent, but it is smaller than that at which dissipation completely
dominates. This certifies that one has to nudge all the scales dominated
by inertial effects in order to have a complete synchronization of the
nudged flow with respect to the reference data. It is important to note
that the Reynolds number of RUN2 is quite high, specially compared to
the standard simulations done in other studies of Data Assimilation.

\begin{figure}
    \centering
    \includegraphics[width=8.5cm]{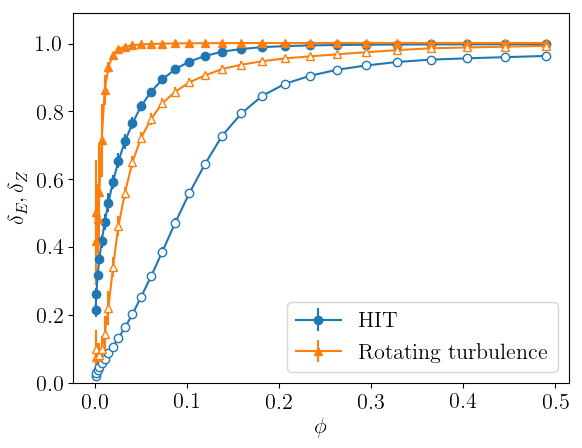}
    \caption{Values of the velocity correlations $\delta_E$, and the
    vorticity correlations $\delta_Z$ (empty symbols) as a function of
    the volume fraction $\phi$ for homogeneous and isotropic turbulence
    (HIT) or under rotation. The HIT simulations have $\alpha t_\eta =
    0.42$ and $\tau/t_\eta=25$, and the rotating turbulnce ones have
    $\alpha t_\eta = 0.3$, and $\tau/t_\eta=33$.}
    \label{rot_scan}
\end{figure}

\subsection{Nudging under the presence of large scale structures} 
\label{large_scale_structures}

Finally, we show the results of nudging a system where large scale
structures are present. As we mentioned in Sec.~\ref{conf_results}, it
is reasonable to expect that different systems can show different
sensitivity to a given nudging scheme. Homogeneous and isotropic
turbulence can be considered  the worst case scenario  as it lacks large
scale coherent structures. In order to show that nudging can indeed be
more efficient in the presence of some coherency into the system, we
applied it to a rotating turbulent flow. Rotating turbulence is known
for generating large columnar vortices with a strong translational
symmetry in the direction parallel to the rotation axis
\cite{Davidson,Sen12,Campagne15,Biferale16}. It is known that nudging
can reconstruct the inverse cascade present in rotating flows
\cite{Clark18}, although this was shown only for spectral nudging.
Equations~\eqref{nsref} and \eqref{nsnudged} were modified by adding a
Coriolis term of the form $-2\Omega \hat{{\bm z}} \times \bu$, with
$\Omega$ being the rotation frequency (see caption of Fig. \ref{rot_viz}
for more details on the simulations)). Figure~\ref{rot_viz} shows a
visualization of horizontal slices of the energy  of the full reference
field, of the filtered/nudging field, and of the nudged/reconstructed
field with applications of the protocol on the configuration domain. The
aforementioned large scale structures are quite easy to spot, and it is
evident how nudging works better in this scenario.
Figure~\ref{rot_scan} compares the value of $\delta_E$ and $\delta_Z$
for one of the previous cases with nudging homogeneous isotropic
turbulence  and one case of nudging rotating turbulence. When the flow
is under rotation, nudging is able to synchronize both the velocity and
vorticity fields to the reference data at much lower volume fractions.
This indicates that nudging can be a very powerful tool in problems that
have large scale structures but are still nonlinear and chaotic.

\section{Conclusions} 
\label{conclusions} 

We have  presented the first systematic application of nudging to three
dimensional homogeneous and isotropic turbulence for big-data
assimilation (high Reynolds number regime). We have investigated the
transition to {\it full} or {\it scale-by-scale} synchronization at
changing the {\it quantity} and the {\it quality} (type) of information
used. In particular, we have implemented nudging with  measurements of
(i) field values on a fixed number of spatial locations (Eulerian case),
(ii) Fourier coefficients of the fields on a fixed range of wavenumbers
(Fourier case), or (iii) field values along a set of moving probes
inside the flow (Lagrangian case). Concerning the {\it quantity} of
information we have shown that full synchronization is achieved as soon
as the $\#_{dof}$ supplied by the nudging field covers a range of scales
that is about one quarter of the dissipative Kolmogorov  wavenumber
(i.e. the largest wavenumber where non-linear inertial
degrees-of-freedom are still active), coinciding with the scale at where
inertial and viscous fluxes match each other.  We have tested this at
both moderate and high Reynolds numbers, where $k_\eta \sim k_0
Re^{3/4}$, and $k_0$ is the energy containing scale.  Similarly for
nudging in configuration space, the critical volume fraction to reach
synchronization is $\phi_c \sim 0.2$. We found that nudging in Fourier
space improves data reconstruction, although paying the price that is
more difficult to apply in realistic field-data applications. Concerning
the {\it quality} of information we found that inputting Lagrangian data
tends to deteriorate the ability to reconstruct but opens a much more
flexible tools for environmental applications. It is also important to
note that the fields be reconstruct have many points (in the order of
$10^7$), so even at high volume fractions, applying a smooth three
dimensional interpolation scheme in order to try to reconstruct the
fields could be prohibtely expensive. Finally, we applied nudging to a
turbulent rotating flow, we showed that despite the dynamics being
richer with a split forward and backward energy cascade
\cite{Alexakis18,Davidson,Pope}, the presence of large scale coherent
structures helps nudging  to reconstruct the reference flow at lower
volume fractions than in the isotropic case, an important fact for many
potential applications.

It is important to remark that our implementation of nudging is
different from the usual one, because we do not supply information about
the external forcing mechanisms in the nudge/reconstructed field
evolution. This is done on purpose, to broaden its applicability to
realistic conditions that are often encountered in the labs or in the
open fields. Furthermore, the application of nudging to big-data goes
well beyond the data-assimilation scope, as it can be seen as an
unbiased equation-informed  tool for classification of complex fields
\cite{Clark18} and/or as a tool to highlight hierarchy of correlations
inside fluid turbulent applications, thanks to the mapping from input to
output  data mediated by the equations of motion. For example, it is
tempting to imagine that nudging could be used in thermal
Rayleigh-B\'enard convection and in MHD to understand the casual
correlation between temperature or magnetic fields with the velocity
field and in bounded flows to disentangle the relative importance of
near-wall regions wrt to bulk for driving the scale and location
dependent turbulent fluctuations. Work in this direction will be
reported elsewhere. 

\begin{acknowledgments}
The authors acknowledge partial funding from the European Research
Council under the European Community's Seventh Framework Program, ERC
Grant Agreement No.~339032. The authors acknowledge help from Michele
Buzzicotti with the Lagrangian simulations.  We acknowledge Francesco Borra,  Massimo Cencini,  Nathan Glatt-Holtz, Cecilia Freire Mondaini, Angelo Vulpiani, Mengze Wang, and Tamer Zaki for
useful discussions.
\end{acknowledgments}

\bibliography{ms}

\end{document}